\documentclass[aps,pra,twocolumn, noshowpacs, groupedaddress,nopacs,floatfix]{revtex4}

\usepackage{graphicx}
\usepackage{xcolor}
\usepackage{amssymb}
\usepackage{amsmath}
\usepackage{times}
\usepackage[colorlinks=true,
            linkcolor = blue,
            urlcolor  = blue,
            citecolor = blue,
            anchorcolor = blue]{hyperref}

\newcommand{\siot}{SiO\textsubscript{2}}
\newcommand{\sitnf}{Si\textsubscript{3}N\textsubscript{4}}
\newcommand{\chft}{CHF\textsubscript{3}}
\newcommand{\ot}{O\textsubscript{2}}
\newcommand{\nt}{N\textsubscript{2}}

\begin{document}
\title{{\fontfamily{ptm}\selectfont \Large Trapping single atoms on a nanophotonic circuit with configurable tweezer lattices}}
\author{May E. Kim$^{1,\dag,\ddagger}$\footnotetext{\small $^{\dag}$ Current address: National Institute of Standards and Technology, 325 Broadway, Boulder, CO 80305.}, Tzu-Han Chang$^{ 1,\ddagger }$,\footnotetext{\small $^{\ddagger}$ These authors contributed equally to this research.} Brian M. Fields$^1$, Cheng-An Chen$^1$, and Chen-Lung Hung$^{1, 2, 3 ,\ast}$\footnotetext{\small $^{\ast}$ e-mail: clhung@purdue.edu.}}
\address{$^1$ Department of Physics and Astronomy, Purdue University, West Lafayette, IN 47907}
\address{$^2$ Purdue Quantum Center, Purdue University, West Lafayette, IN 47907}
\address{$^3$ Birck Nanotechnology Center, Purdue University, West Lafayette, IN 47907}

\begin{abstract}
%\section{Abstract}
Trapped atoms near nanophotonics form an exciting platform for bottom-up synthesis of strongly interacting quantum matter. The ability to induce tunable long-range atom-atom interactions with photons presents an opportunity to explore many-body physics and quantum optics. Here, by implementing a configurable optical tweezer array over a planar photonic circuit tailored for cold atom integration and control, we report trapping and high-fidelity imaging of one or more atoms in an array directly on a photonic structure. Using an optical conveyor belt formed by a moving optical lattice within a tweezer potential, we show that single atoms can be transported from a reservoir into close proximity of a photonic interface, potentially allowing for the synthesis of a defect-free atom-nanophotonic hybrid lattice. Our experimental platform can be integrated with generic planar photonic waveguides and resonators, promising a pathway towards on-chip many-body quantum optics and applications in quantum technology.
\end{abstract}
\maketitle

\section{Introduction}
Coupling an array of trapped atoms to an engineered photonic environment opens up more regimes in quantum optics \cite{Chang_2014_photon} and many-body physics \cite{GeorgescuRMP2014,douglas_quantum_2015,gonzalez-tudela_subwavelength_2015,ChangRMP2018}. Integrating cold atoms with nanophotonic platforms has so far been restricted to discrete, suspended structures of quasi-linear geometry \cite{vetsch_optical_2010,goban_demonstration_2012,mitsch_quantum_2014,kato_strong_2015,sorensen_coherent_2016,corzo_large_2016,Solano_longrange_2017,goban_superradiance_2015,hood2016atom,thompson_coupling_2013,tiecke_nanophotonic_2014} due to the requirement of open optical access for laser cooling and loading of cold atoms from freespace. Single atom manipulation and direct imaging on nanostructures also remains elusive. Beyond these technical challenges, there is strong motivation to migrate cold atoms to planar photonic platforms, which may offer a wide variety of quantum functionalities with increased dimensionality and scalability. Planar structures, such as two-dimensional (2D) photonic crystals \cite{Majumdar2DPhC2012,gonzalez-tudela_subwavelength_2015,yu2018phc} or coupled resonator optical waveguides \cite{Hafezi_qh_2013}, can induce coupling between atoms and photons with engineered chiral quantum transport \cite{lodahl_chiral_2016} and non-isotropic interactions \cite{gonzalez_anisotropic_2017}, also making it possible to explore topological physics \cite{Hafezi_qh_2013,hung_quantum_2016,Khanikaev_2d_topological_2017,Noh_Many-body_light_2017} or vacuum induced quantum phase transitions \cite{gonzalez-tudela_subwavelength_2015}. Realizing these remarkable possibilities requires a robust experimental scheme and an enabling photonic platform for efficient loading and interfacing with cold atoms. 

Recent development of optical tweezer cold atom assemblers \cite{lester_rapid_2015,endres_atom-by-atom_2016,barredo_atom-by-atom_2016} provides invaluable toolbox for synthesizing atom-nanophotonics hybrid quantum matter. While guided modes in nanophotonics can be utilized for global evanescent-wave trapping \cite{le_kien_atom_2004,hung_trapped_2013} and inducing cooperative atom-photon coupling \cite{hood2016atom}, implementing independent control using optical tweezer trapping \cite{thompson_coupling_2013}, manipulation, and single atom imaging techniques offers a complementary toolkit for arbitrary state preparation, local addressing, and site-resolved final state detection.

In this article, we report single atom trapping and direct imaging on a planar photonic circuit in a configurable tweezer array. We demonstrate that single cesium atoms can be loaded into an optical tweezer that is tightly focused on the surface of a nanostructure. These trapped atoms can be fluorescence imaged on an electron-multiplied charged coupled device (CCD) camera, through the same objective that is utilized to project the tweezer beam. A tweezer beam reflected from a planar structure forms an inhomogeneous lattice of micro-traps that can localize multiple cold atoms. We show that such a tweezer lattice can be converted into an optical conveyor belt, transporting trapped atoms into or out of the tweezer focus for vertical positioning near the planar dielectrics for atom-nanophotonics lattice assembly. 
\section{Results}
%\textbf{Planar photonics circuit design and experiment scheme.} 

\subsection{Imaging single atoms on a nanophotonic membrane circuit}
As illustrated in Fig. 1, we have designed and fabricated a transparent, optically flat photonic membrane, consisting of a 2~$\mu$m thick SiO$_2$ layer and a 550~nm thick nitride (Si$_3$N$_4$) bottom-layer with high tensile stress after releasing from a silicon substrate within a 2~mm $\times$ 8~mm transparent window~\cite{appendix}. The transparent membrane allows full optical access for laser cooling and optical control of cold atoms. Discrete or coupled arrays of photonic structures, such as ring/racetrack resonators \cite{ritter_ring_2016} or coupled resonator optical waveguides \cite{Morichetti2012}, among general planar structures, can be patterned in an additional nitride top-layer on the membrane to induce atom-light interactions for designer quantum functionalities ~\cite{appendix}. 

\begin{figure*}[t]
\centering
\includegraphics[width=1.5\columnwidth]{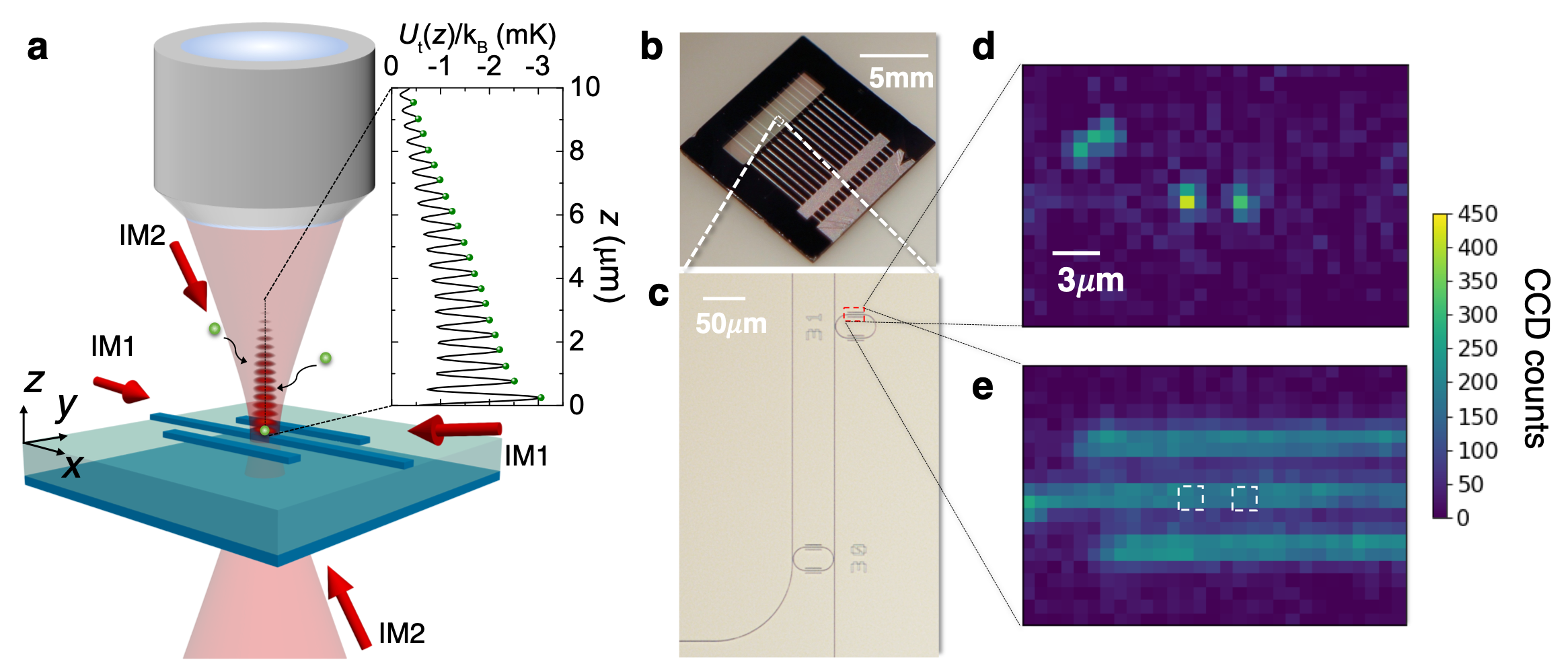}
\caption{\textbf{Single atom trapping and imaging on a planar dielectric nanostructure.} \textbf{a}, Simplified schematic of the microscope objective (numerical aperture NA$=0.35$) projecting a tightly focused optical tweezer beam ($1/e^2$ beam waist $1.2~\mu$m) onto a planar photonic structure. Light and dark shaded rectangles represent SiO$_2$ and Si$_3$N$_4$ dielectric structures, respectively. Due to finite surface reflectance, a lattice of micro-traps (dark red shaded region) forms within the tweezer potential. Inset shows the potential line-cut $U_\mathrm{t}(z)$ through the center of the tweezer lattice. Single atoms (green spheres) are cooled and loaded into the traps and are fluorescence imaged while scattering photons from a pair of counter-propagating, near-resonant beams (red arrows marked by either IM1 or IM2). \textbf{b}, The photonic membrane circuit. A zoom-in view within the dashed box is shown in \textbf{c}, where sample planar structures in the top-layer of the membrane are visible.  \textbf{d}, Atomic fluorescence image (using IM1) of two adjacent, loaded tweezer traps focused in the red box region as illustrated in \textbf{c}. \textbf{e}, Bright field image of the nanostructure recorded under the identical image focus and field of view ~\cite{appendix}. Dashed boxes mark the location of the atoms in \textbf{d}. Pixel size:~(800~nm)$^2$.}
\label{fig1}%
\end{figure*}

We project an array of tweezer beams on top of the membrane through the control of a pair of acousto-optic deflectors (AODs). A stationary lattice of micro-traps (Fig.~\ref{fig1}a) forms within individual tweezer potentials. The closest site is $\sim 200~$nm above the surface, well within the evanescent-wave range of a guided mode at the atomic resonance ($z<\lambda_\mathrm{a}=852~$nm), and is stable against atom-surface Casimir-Polder interactions~\cite{appendix}. To fill the tweezer lattices, a magneto-optical trap first guides a cold cloud of cesium atoms into close proximity of the membrane surface, followed by polarization-gradient cooling (PGC). Typical atom number density is $\rho_0 \approx 3.5\times10^{9}~$ cm$^{-3}$ near the surface with a temperature $T\approx15~\mu$K. During PGC, the tweezer beams are ramped on to full power (5~mW) to form deep micro-traps, up to $|U_\mathrm{t}| \approx k_\mathrm{B} \times 3~$mK, near the structure surface (Fig.~\ref{fig1}). To achieve uninterrupted laser cooling in these deep traps, we adopt a magic wavelength $\lambda_\mathrm{t}=$935~nm for the optical tweezers to eliminate differential light shift in the cooling transition \cite{Nicholas2017}. Following 10~ms of PGC in tweezers, the cooling beams are extinguished for at least 50~ms, ensuring that unbound atoms can permanently leave the trap region. We then turn on a pair of linearly-polarized, near-resonant beams for 30~ms to record atomic fluorescence on an CCD camera. 

Figure~\ref{fig1}d shows the single-shot fluorescence image of two loaded tweezer traps, which manifest as localized bright spots with high fluorescence counts. In this example, both tweezer beams are aligned to a linear structure (an optical waveguide) of 870~nm width, and are separated in-plane by $\Delta x = 3~\mu$m. The image focus is on the structure, which is nearly dark due to polarization filtering~\cite{appendix}. The measured atomic fluorescence spot size ($1/e^2$ radius $\lesssim1.5~\mu$m) indicates that trapped atoms are well within the depth of field ($z \lesssim 10~\mu$m ) from the structure surface.

\begin{figure}[t]
\centering
\includegraphics[width=1\columnwidth]{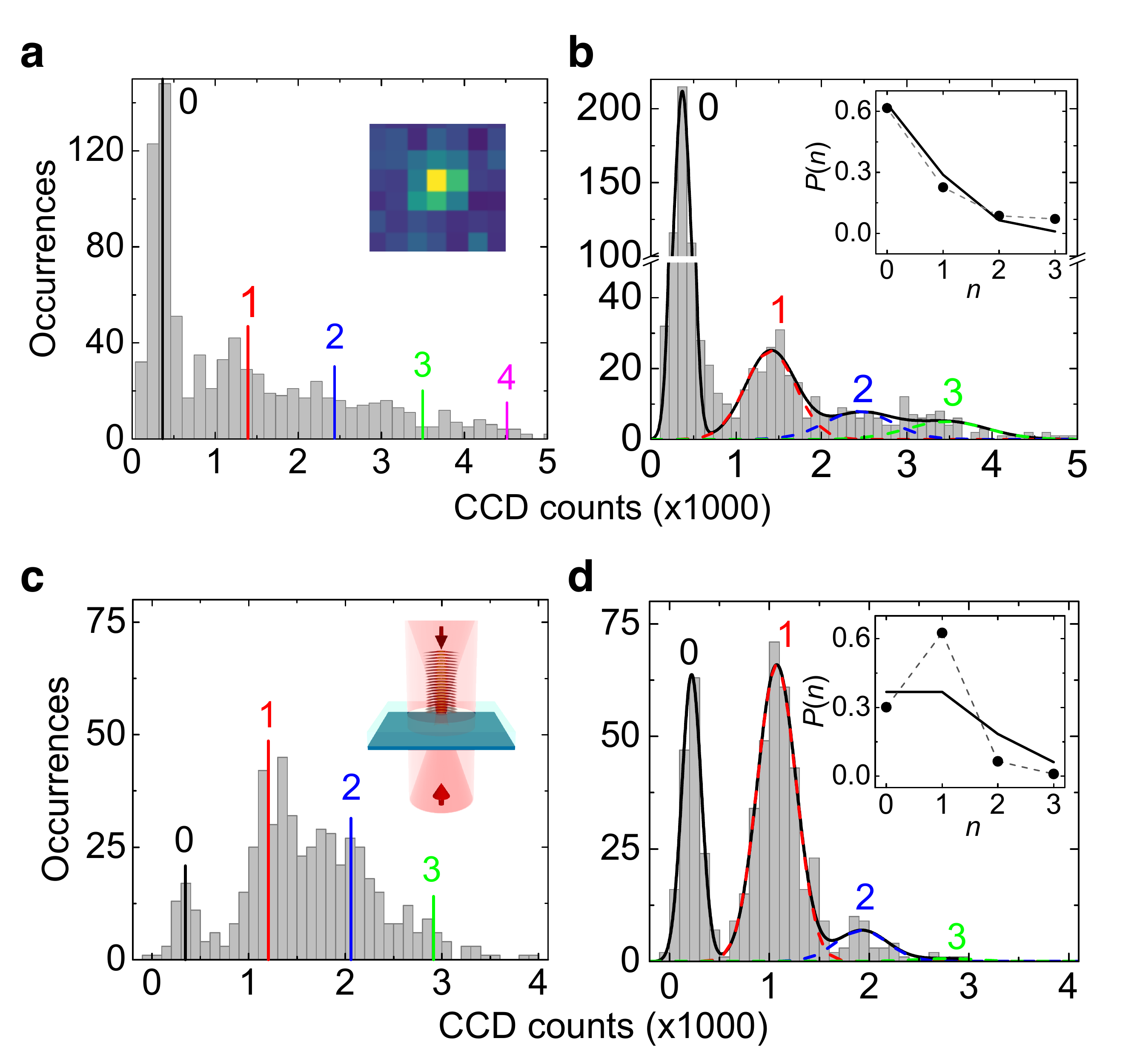}
\caption{\textbf{Single atom loading and fluorescence counting on a nanostructure.} \textbf{a} and \textbf{b}, Histograms of fluorescence counts within a single tweezer trap (inset) on the nanostructure as shown in Fig.~\ref{fig1}, recorded in two consecutive imaging periods that are separated by a 40~ms dark time. \textbf{c} and \textbf{d}, Histograms of fluorescence counts for trapping in a tweezer lattice on a membrane, recorded in the two imaging periods. Here, a phase coherent, counter-propagating dipole beam is added to form an optical lattice within the treezer trap (inset); for potential curves, see Fig.~\ref{fig:conveyor}. Single atom fluorescence peaks in both cases manifest in the second imaging period.  In \textbf{b} and \textbf{d}, composite Gaussian fits (black solid curves) indicate the count distributions of $n=0,1, 2, ...$ trapped atoms (color dashed curves) and their averaged count contributions (vertical lines in \textbf{a} and \textbf{c}). Inset in \textbf{b} (\textbf{d}) shows the fitted probability $P(n)$ of trapping $n$ atoms (circles) in a tweezer lattice and a Poisson fit (solid lines) with $\bar{n}=0.45$ ($\bar{n}=1$).}
\label{fig:wg}%
\end{figure}

Fluorescence counts collected within a single tweezer trap are analyzed for loading statistics. Figure~\ref{fig:wg}a shows a histogram of counts from more than 800 experiment repetitions. Around 60\% of the time, we observe atomic fluorescence which is distinct from the background signal (sharp peak in Fig.~\ref{fig:wg}a). However, atom number-resolved peaks cannot be identified. As suggested by a Monte Carlo simulation~\cite{appendix}, around $10$ deepest sites (within $5~\mu$m of focus) in the tweezer beam can stably trap atoms. Away from the focus, loosely bound atoms with reduced fluorescence counts, if present, may smear out otherwise number-resolved signals.

To distinguish tightly trapped atoms, we initiate a second imaging period 40~ms after the first imaging pulse terminates. In Fig.~\ref{fig:wg}b we indeed observe signature of single atom fluorescence from those atoms surviving the first imaging procedure. We find around 40\% probability of trap population $n\geq 1$. The probability distribution is consistent with a Poisson statistics of $\bar{n}=0.45$. Using the fitted single atom count from Fig.~\ref{fig:wg}b and the histogram in Fig.~\ref{fig:wg}a, we estimate that $\bar{n}_\mathrm{wg} \gtrsim 1$ atoms are loaded into the trap. From the Monte Carlo simulations performed~\cite{appendix}, we expect the first trap site closest to the dielectric surface to be filled $\sim$2\% of the time when an atom is localized within the tweezer trap.

\begin{figure}[t]                                                                                       
\includegraphics[width=1.\columnwidth]{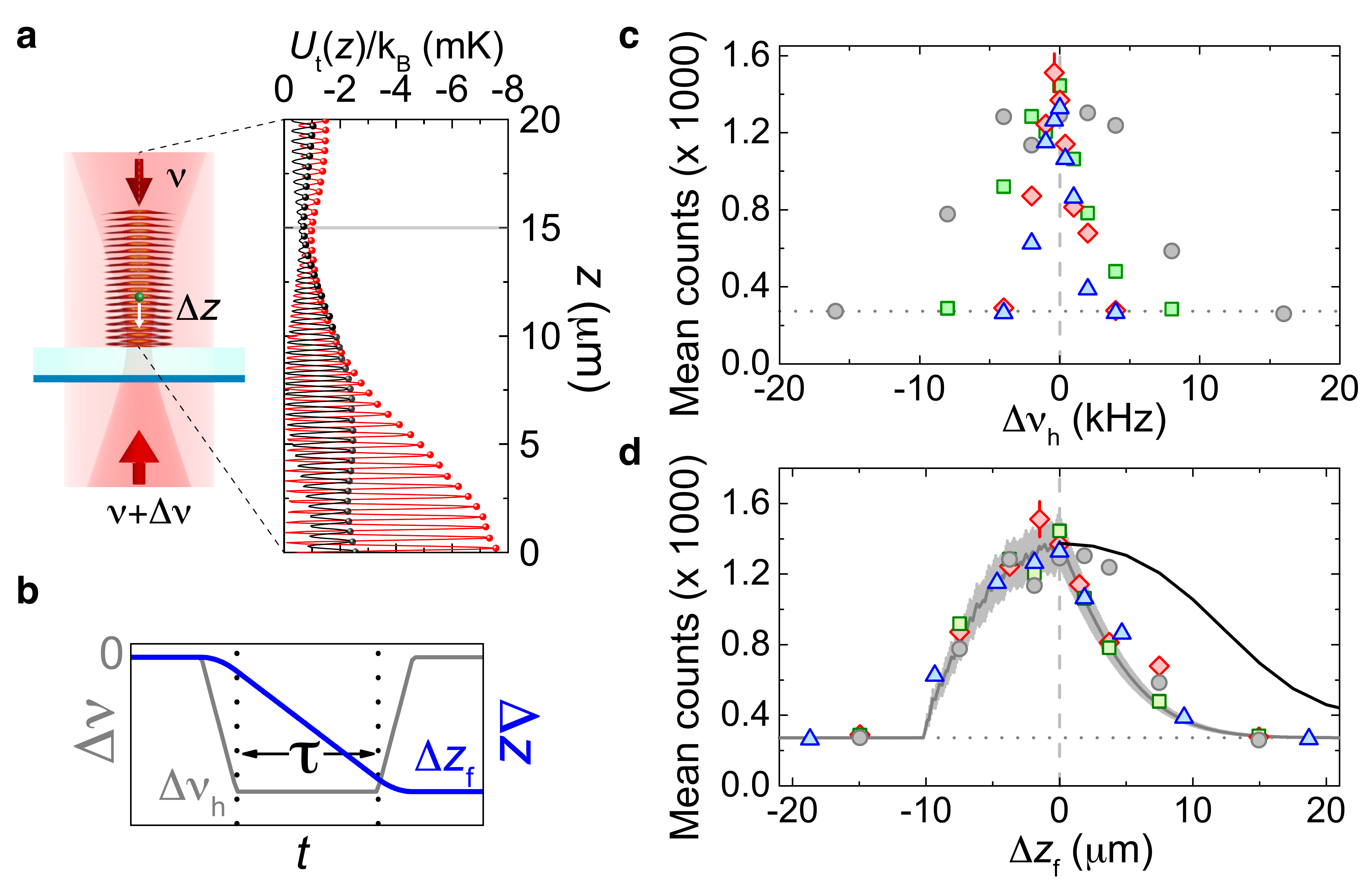}%  
\caption{\textbf{Optical conveyor belt in a tweezer lattice.}  \textbf{a}, Schematic and potential line-cuts $U_\mathrm{t}(z)$ of the optical conveyor belt, shown at instances when the back dipole beam is in-phase (red curve) or out-of-phase (black curve) with respect to the tweezer beam surface reflection. The lattice site centers are marked by the color spheres. Note that the trap depth near the surface is large $|U_\mathrm{t}|>k_\mathrm{B} \times 2~$mK until near the depth of field $z_\mathrm{dof} \approx 15~\mu$m marked by the gray line. \textbf{b}, Transport distance $\Delta z$ (blue curve) controlled by the frequency detuning $\Delta\nu$ (gray curve) of the bottom dipole beam, which is linearly ramped to $\Delta \nu_\mathrm{h}$ in 1~ms, held for a time $\tau$, and then ramped back to zero. \textbf{c}, Mean fluorescence counts versus the detuning $\Delta\nu_\mathrm{h}$ at various hold times $\tau$=1 (gray circles), 3 (green squares), 7 (red diamonds), and 9 (blue triangles)~ms. \textbf{d}, The same sets of measurements (filled symbols) plotted against the final distance $\Delta z_\mathrm{f}$. Black solid line at $\Delta z_\mathrm{f} >0$ is a model considering count reduction only due to defocusing. Gray solid line is an empirical fit and the shaded region takes into account the error of the mean; see text. The horizontal dotted lines mark the mean background counts. Error bar represents standard deviation of the mean.}
\label{fig:conveyor}
\end{figure} 

\subsection{Trap loading and transport in an optical tweezer lattice}

To further improve trap loading efficiency, we introduce a phase coherent, counter-propagating optical beam of a larger beam waist (7~$\mu$m) to increase the trap volume and also form a stronger tweezer lattice (Fig.~\ref{fig:wg}c,d). The beam is sent from the bottom side of the transparent window and is ramped up to a transmitted power of 84~mW simultaneously with the top tweezer beam. To keep the discussion general, from here forward we discuss loading directly on the membrane without additional nanostructure in the top layer so one can assume the bottom dipole beam profile is smooth above the membrane. Fluorescence imaging is performed using a pair of beams (IM2) as shown in Fig.~\ref{fig1}~\cite{appendix}.

As shown in Fig.~\ref{fig:wg}c, trap loading probability increases up to 90~\% with $n\geq1$. Single atom counting statistics again manifests in the second imaging period in Fig.~\ref{fig:wg}d, where a prominent single atom peak indicates that around 60~\% of the time an atom is tightly trapped within the tweezer. The probability distribution is clearly sub-Poissonian, with $\langle \delta n^2 \rangle = 0.35<\bar{n}=0.77$, likely due to collisional blockade effect within the dipole trap \cite{Schlosser2002} before the atoms are cooled into individual micro-traps. The estimated trapped atom number is $\bar{n}_\mathrm{lattice} \gtrsim 1.6$.

Introducing a phase-coherent bottom dipole beam also allows the control of atom position within the tweezer. The stationary tweezer lattice can be overridden by interfering with a counter-propagating beam with stronger intensity than that of the reflected tweezer beam from the membrane. By controlling the optical phase difference between the two beams, we can transport a trapped atom within a tweezer like a conveyor belt \cite{Schrader2001,Miroshnychenko2006}, as illustrated in Fig.~\ref{fig:conveyor}a.

Following trap loading and a clean-up procedure~\cite{appendix}, the conveyor transport is initiated by introducing a small frequency detuning $\Delta \nu$ of the bottom beam. As shown in Fig. ~\ref{fig:conveyor}b, the transport distance $\Delta z(t) = \frac{\lambda_\mathrm{t}}{2}\int_0^t \Delta \nu(t') dt'$
is controlled by the detuning $\Delta \nu_\mathrm{h}$ and the hold time $\tau$. Prior to imaging, the bottom dipole beam is quickly ramped off in 2~ms, reverting the tweezer trap back to the stationary lattice configuration. Figure~\ref{fig:conveyor}c plots the measured mean fluorescence counts averaged over multiple experiment repetitions, which decreases significantly at nonzero $|\Delta \nu_\mathrm{h} |$ and at larger $\tau$. These measurements collapse into a single curve when we plot them against final transport distance $\Delta z_\mathrm{f}$, as shown in Fig.~\ref{fig:conveyor}d, suggesting that the count decrease is due to atom transport in the tweezer lattice instead of other loss mechanisms, such as parametric or noise heating, that should separately depend on $|\Delta \nu_\mathrm{h}|$ or hold time $\tau$. 

The count asymmetry in Fig.~\ref{fig:conveyor}d is attributed to the conveyor transport near the membrane surface. With $\Delta z_\mathrm{f}>0$, atoms are transported away from the tweezer focus until they approach $z_\mathrm{dof} \approx 2\lambda_\mathrm{t}/\mathrm{NA}^2 = 15~\mu$m where the tweezer intensity on the optical axis vanishes. The count reduction is primarily due to atoms being heated out of the trap during imaging or escaping during transport, and is much more than what would be suggested by a defocused atom image~\cite{appendix}. Concerning the distance scale involved, $\Delta z_\mathrm{f} \gg \lambda_\mathrm{a}$, the modification of dipole emission rate due to coupling to the membrane guided modes also plays insignificant role. On the other hand, for conveying downward, mean counts gently reduce until $\Delta z_\mathrm{f} \lesssim -10~\mu$m, beyond which no atomic fluorescence can be detected. This results from multiple trapped atoms, randomly distributed along the tweezer lattice, being pulled closer to the membrane surface where the micro-traps are the strongest. These atoms can be imaged well until they are too close to or eventually adsorb on the membrane surface.  

From Fig.~\ref{fig:conveyor}d, we can infer the trap range and atom number. Using the data from $\Delta z_\mathrm{f} \geq 0$ as an empirical model for the fluorescent counts versus trapped atom position, we can fit the data for $\Delta z_\mathrm{f} <0$ by assuming a Poisson average of $\bar{n}_\mathrm{lattice}$ trapped atoms initially distributed along the lattice with random site positions $0<z_i \leq z_\mathrm{max}$ and transport distance $\Delta z_\mathrm{f}$ ~\cite{appendix}. We obtain a reasonable fit with $\bar{n}_\mathrm{lattice}=3.6$ atoms and $z_\mathrm{max} \approx $10$~\mu$m, consistent with the expected trap range. We note that this simple model does not take into account the surprising sub-Poissonian distribution near the tweezer focus, as seen in Fig.~\ref{fig:wg}d.

\subsection{Experimental outlook}

Conveyor-belt transport of an atom within a tweezer trap can be utilized to improve the loading probability of the trap site closest to the dielectric surface, starting from a random initial vertical position and stopping at the first trap site before the atom hits the surface.  This can be implemented on a nanostructure with a coupled resonant probe to feedback-control the operation. Through monitoring the guided probe transmission in realtime, the presence of a single atom in the evanescence region ($z<\lambda_\mathrm{a}$) can be inferred by detecting a significant drop in probe transmission due to strong atom-light interactions. Up to 70~\% drop is expected when an atom is coupled to the band edge mode of a photonic crystal waveguide as recently demonstrated in an experiment \cite{goban_superradiance_2015}. Recently, a clocked delivery of cold atoms to the same waveguide has been reported in Ref. \cite{burgers2019clocked} using an optical conveyor-belt. Similar feedback can also be achieved by coupling to a resonator of high quality factor and moderate mode volume \cite{ChangRMP2018}. 

\begin{figure}[t]
\includegraphics[width=0.8\columnwidth]{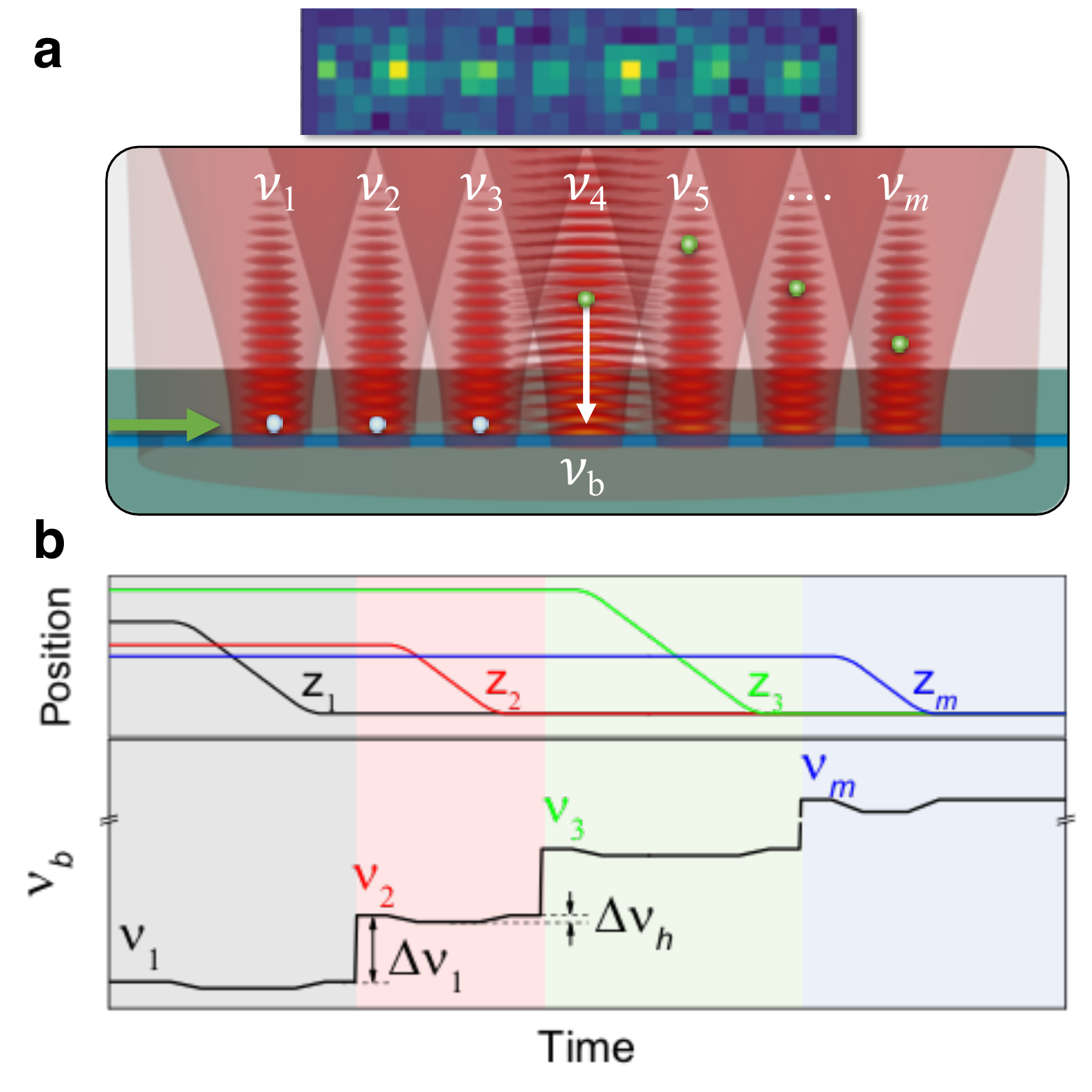}%  
\caption{\textbf{Scheme for conveyor belt assisted assembly of an atom array on a nanophotonic waveguide.} \textbf{a}, Assembly begins with an array of tweezers each controlled by a frequency tone $\nu_i$ and filled with $n\geq1$ trapped atoms as shown in the single-shot fluorescence image (inset), followed by the conveyor transport in individual tweezers as schematically shown, with feedback control by monitoring the transmission of a resonant guided probe (green arrow) for detecting single atoms (green spheres) and pumping them into a dark state (blue spheres). \textbf{b}, Schematic frequency-tuning profile of the bottom dipole beam ($\nu_b$) to control the atom position $z_i$ in individual tweezers. When the frequency shift $|\nu_\mathrm{b} - \nu_i| \gg f_\mathrm{a}$ (axial trap frequency), tweezer $i$ forms a time-averaged stationary lattice as shown in Fig.~\ref{fig1}a. The time-dependent conveyor transport can be initiated only when $ |\nu_\mathrm{b} - \nu_i| \ll f_\mathrm{a}$, as illustrated in \textbf{a} where $\nu_\mathrm{b} \approx \nu_4$. }
\label{fig:assembly}
\end{figure}

This conveyor belt technique can be scaled up to an array configuration by beginning with a filled tweezer array as shown in Fig.~\ref{fig:assembly}a, where each trap is controlled by one frequency tone $\nu_i$ in the AOD ($i=1,2,...,m$). For the tone spacing much greater than the axial trap frequency $\Delta \nu_i = |\nu_i-\nu_{i+1}| \gg f_\mathrm{a}$, where typically the axial trap frequency $f_\mathrm{a} <900~$kHz ~\cite{appendix}, one counter-propagating dipole beam that spatially overlaps with all tweezers can transport trapped atoms in individual tweezer lattices, one at a time. Figure~\ref{fig:assembly}b schematically illustrates a profile of the frequency $\nu_\mathrm{b}$ of the (counter propagating) bottom dipole beam for such an operation. Within time segment $i$ marked by a color shaded area, the conveyor transport initiates when $\nu_\mathrm{b}\approx\nu_i$ and can be terminated by feedback from a resonant probe, followed by a rapid change (within a time $\ll 1/f_\mathrm{a}$) to the next frequency tone $\nu_\mathrm{b}=\nu_{i+1}$ to convert the $(i+1)$-th tweezer lattice into a conveyor belt and revert the $i$-th tweezer back to the original stationary tweezer lattice. Following the transport, an atom can be optically pumped to a dark state using the guided mode, awaiting further operations. We expect transport in each tweezer lattice should finish within $\tau<5$~ms. With our measured tweezer trap lifetime $\gtrsim 900~$ms, tens of trapped atoms may be assembled using an array of tweezer lattices and a bottom dipole beam of a moderate power.

\section{Discussion}
In summary, using a configurable tweezer lattice, we show that a single or an array of atoms can be loaded, transported, and imaged on a planar photonic circuit. We further propose that conveyor-belt transport can be utilized to assemble an atom array on a nanophotonic waveguide. Our experimental platform and technique extend beyond existing demonstrations of trapped atoms on suspended, quasi-linear nanophotonics such as nanofibers \cite{vetsch_optical_2010,goban_demonstration_2012,kato_strong_2015,sorensen_coherent_2016,corzo_large_2016}, photonic crystal waveguides \cite{goban_superradiance_2015,hood2016atom}, and cavities \cite{thompson_coupling_2013,tiecke_nanophotonic_2014}, opening up more possibilities of coupling trapped atoms to lithographic planar photonic structures with broad applications and quantum functionality \cite{yu2018phc,gonzalez-tudela_subwavelength_2015}. Our ability to perform single atom fluorescence imaging on a dielectric surface allows for state-sensitive, atom-resolved detection that is complementary to conventional guided mode probing techniques. Lastly, our photonics membrane platform can be readily extended to include light-coupled high-quality resonator waveguides, enabling future studies of many-body quantum optics or even the synthesis of an array of ground state molecules \cite{Rios_molecule_2017}.

\section*{Acknowledgements}
We are grateful for the support and discussions from H. J. Kimble. We thank S.-P. Yu, and Y. Xuan for discussions. We acknowledge T.-W. Hsu, B. Edelman, W. Wang, J. Quirk, and K. Knox for technical assistance. Funding is provided by the AFOSR YIP (Grant NO. FA9550-17-1-0298), ONR (Grant NO. N00014-17-1-2289) and the Kirk Endowment Exploratory Research Recharge Grant from the Birck Nanotechnology Center.

\bibliography{imaging_arXiv}
\renewcommand\thefigure{A\arabic{figure}}    
\setcounter{figure}{0} 
\appendix
\section{ The vacuum system} 
\begin{figure}[b]
\includegraphics[width=0.8\columnwidth]{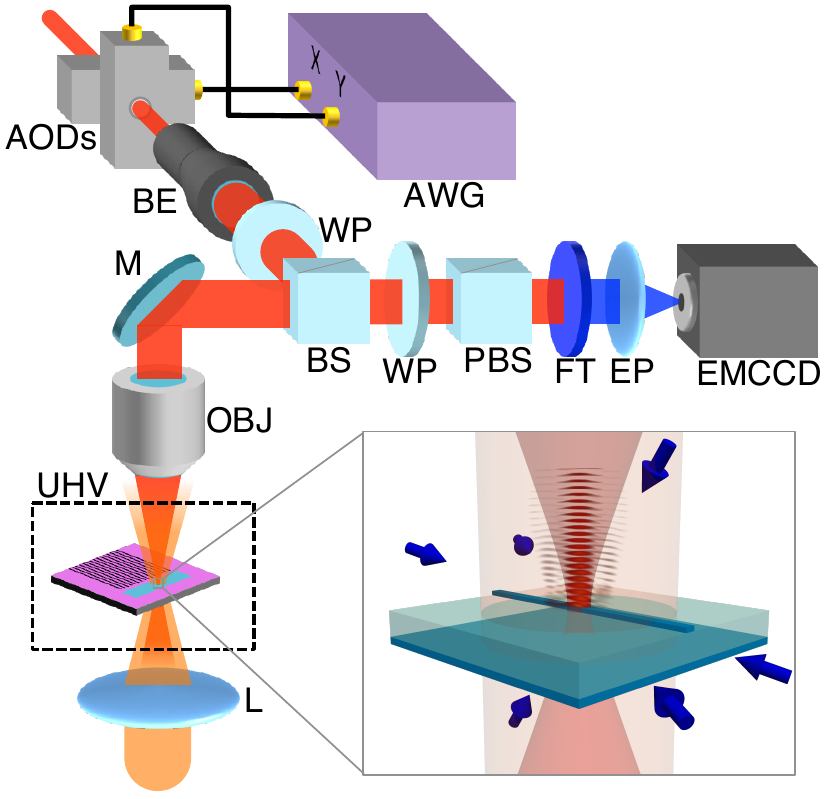}%  
\caption{\textbf{Schematic of the experiment apparatus.} Cold atoms are prepared on top of a photonic circuit in an ultrahigh vacuum chamber (dashed box marked by UHV) with three retro-reflected cooling beams (blue arrows) forming a magneto-optical trap. The tweezer beam is steered by two acousto-optic deflectors (AODs) that are driven by an arbitrary waveform generator (AWG), followed by beam expansion (BE) and projection through the microscope objective (OBJ). The bottom dipole beam is projected from the bottom of the membrane through a lens (L). Atomic fluorescence is filtered through waveplates (WP), a polarization beam splitter (PBS) and stacked interference filters (FT), and is recorded on an electron multiplying charge-coupled device (EMCCD) through an eyepiece (EP) that can independently adjust the image focal plane. Fluorescence imaging beams are shown in Fig. 1 and are not drawn here.}
\label{figSM_app}
\end{figure}

Our optical chip (Figs.~\ref{fig1} and \ref{figSM_app}) is glued onto a vacuum compatible holder, which is docked on a linear-translation and rotation stage inside the ultrahigh vacuum (UHV) chamber for positioning. Although not discussed in this study, optical fibers can be glued onto on-chip fiber grooves for coupling light to waveguide buses. These fibers are guided outside the chamber via vacuum feedthroughs, and are reserved for future studies of atom-light interactions. 

\section{ Chip fabrication process} 
Our chip fabrication begins with low pressure chemical vapor deposition (LPCVD) followed by high-temperature annealing to first grow a 550~nm thick \sitnf~bottom-layer film and then a 2~$\mu$m thick \siot~mid-layer film on a 4 inch silicon wafer. The annealing temperatures are around 1100 and 900 $^\circ$C for the \sitnf~and \siot~layers, respectively. The chosen \siot-\sitnf~thickness and the annealing result in a positive tensile stress around 100 MPa after the \siot-\sitnf~membrane is released from the silicon substrate, keeping the suspended membrane stressed and optically flat (measured flatness $<200~$nm within a 1~mm $\times$ 1~mm area). Additional film in the top layer, 360~nm thick \sitnf~for this study, is LPCVD grown, followed by dicing the wafer into 12 mm $\times$ 12 mm chips.

To pattern additional nanostructures in the top layer, electron beam lithography (EBL) with MaN 2403 negative tone resist is performed on a 100 KeV EBL system (JEOL JBX-8100FS). After e-beam exposure and resist development, device pattern is transferred onto the top \sitnf~layer with an inductively-coupled reactive-ion etching (ICP-RIE) tool (Panasonic E620). In the dry etching process, a gas mixture of \chft /\ot /\nt~is employed to achieve high selectivity to the \siot~layer and low sidewall roughness.

To release the membrane from the substrate, a back window is patterned by photo-lithography, and is dry-etched using ICP-RIE (Panasonic E620) and deep RIE (STS-ASE) tools until the remaining silicon substrate is $10\sim20\mu$m thick. The final membrane release process is performed using low temperature TMAH wet etching. On the front side of the chip, a set of U-shape grooves for optical fiber edge coupling, visible in Fig.~\ref{fig1}b near the edge of the transparent window, can be patterned using similar procedures prior to etching the back window. Throughout the window-release process, a PMMA resist is applied to coat and protect the front side of the chip. 

Figure~\ref{fig1}b-c show a sample optical chip, with arrays of resonator waveguides coupled to bus waveguides on the membrane, and with fiber U-grooves for guiding light into the circuits. Detailed optical functionality, characterization and the result of atom-light couplings will be reported elsewhere. 

\section{ Adjusting the tweezer beam focus and the image plane} 
There is finite chromatic aberration presenting in our commercial apochromatic objective (corrected for 3~mm thickness of our vacuum glass viewport). As a result, the tweezer beam focus (at $\lambda_\mathrm{t}=935~$nm) and the image plane (at $\lambda_\mathrm{a}=852~$nm) need to be independently and carefully adjusted to both coincide on the membrane surface. We use the photonic structure as shown in Fig.~\ref{fig1} to assist in the focusing procedure. At $\lambda_\mathrm{t}=935~$nm and at small incidence angles $\theta<20^\circ$, the reflectance of the waveguide (870~nm wide and 360~nm thick) is $R_\mathrm{w} \approx 0.03$, which is smaller than the reflectance of the membrane $R_\mathrm{m} \approx 0.3$. We bring the minimum beam waist of the tweezer beam onto the membrane/nanostructure through minimizing the reflected power. We then adjust the position of the eyepiece (InfiniTube Standard System) to focus the image of the nanostructure taken at $\lambda_\mathrm{a}$ on the EMCCD. The position uncertainty between the tweezer focus and the membrane surface is estimated to be $\delta z<1~\mu$m.

\section{On-chip laser cooling and trap loading} 
\begin{figure}[t]
\includegraphics[width=1\columnwidth]{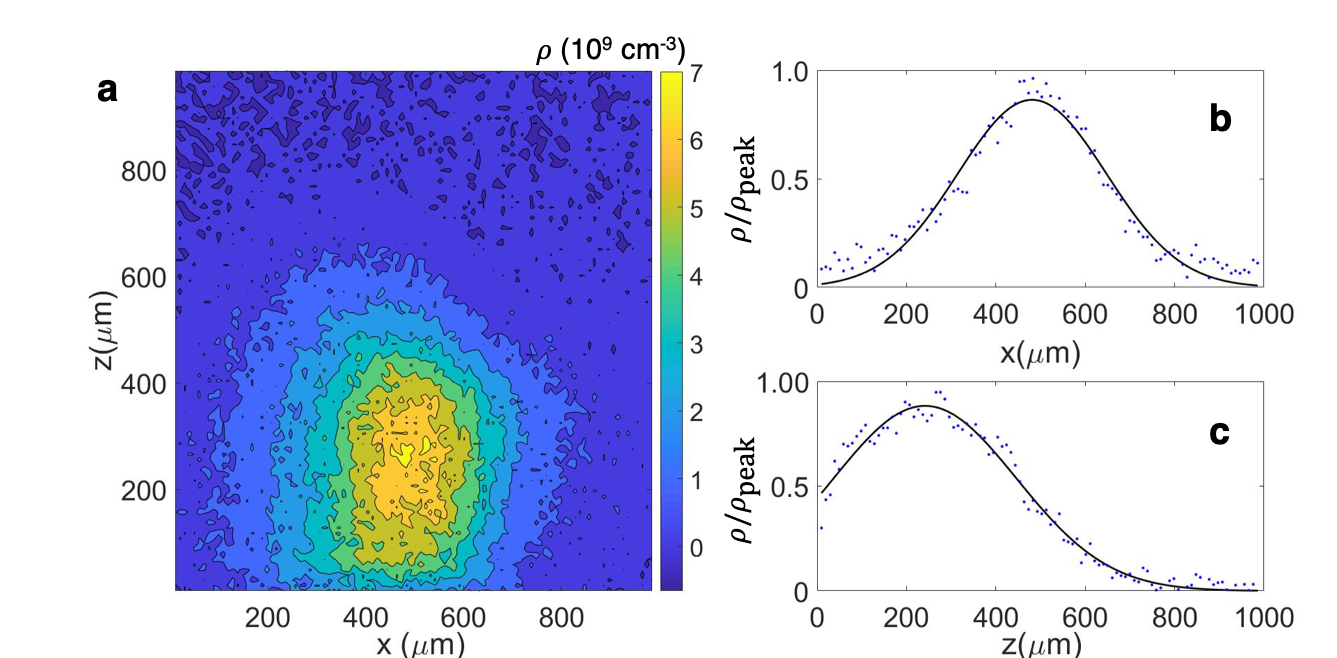}%  
\caption{\textbf{Atom number density near the membrane surface.} (a) On-chip atomic density distribution. Membrane surface is in the $z=0$ plane. (b) and (c), Line-cuts of the density distribution through the cloud center, normalized by the peak density $\rho_{\mathrm{peak}} \approx 7\times10^9$ cm$^{-3}$. Projected atom number density at $z<25~\mu$m near the chip surface is $\rho_0\approx 3.5\times 10^9$ cm$^{-3}$.}
\label{figSMod}
\end{figure}

Our experiment begins with around $10^6$ laser-cooled cesium atoms collected in the vicinity of the optical chip, followed by transporting the atoms onto the membrane using a velocity selective cooling method \cite{Shang_velocity_cooling_1991}. The atoms are then recaptured in a magneto-optical trap (MOT) formed by three circularly polarized, retro-reflected laser beams of $\sim$2~mm beam waist, which intersect directly above the membrane surface; two nearly horizontal beams intersect the chip surface with a 75~degree incidence angle while the third beam crosses the chip with a $\sim$30~degree incidence angle; see Fig.~\ref{figSM_app}. Membrane reflectances at these crossing angles are $R_{\theta=75^\circ}\approx 0.88~(0.24)$ and $R_{\theta=30^\circ}\approx 0.13~(0.09)$, for S-(P-)polarization. All the cooling beams are launched from the bottom side of the membrane and are re-collimated and retro-reflected to balance the radiation pressure. Following MOT, the atoms are then polarization-gradient cooled (PGC) in 10~ms to $T\leq15~\mu$K at a detuning -120~MHz from resonance, forming a cold atom reservoir for this study. The measured $1/e$-lifetime of the PGC atoms on membrane is $\tau_\mathrm{life}\sim 43~$ms. This short life time is partially due to imperfect radiation pressure balancing and also due to atoms adsorbed on the membrane surface. Figure \ref{figSMod} (a) shows a density contour plot deduced from a freespace absorption image. The extrapolated atom number density near the chip surface is $\rho_0\sim 3.5\times10^9~$ cm$^{-3}$.

During the PGC, the tweezer beams (and the bottom dipole beam, if required) are ramped on to full power of 5~mW (84~mW) in 5~ms, followed by additional 10~ms wait time that allows nearby atoms to be cooled into the tweezer trap. We found that trap loading probability becomes significant only when we use the magic wavelength $\lambda_\mathrm{t} = 935~$nm to form the tweezer traps. Trap loading becomes inefficient away from the magic wavelength, even after an exhaustive search for proper detunings and powers of the MOT cooling lasers, repumpers and the imaging beams. This is in sharp contrast to tweezer loading in freespace, which we have tested to easily achieve $>60~$\% single atom loading efficiency in a range of trap wavelengths $\lambda_\mathrm{t} = 860\sim910~$nm (but with different MOT and imaging settings). This indicates that uninterrupted cooling is very important for cooling atoms into tight traps near a planar dielectric surface.

Following PGC, the cooling light is extinguished for at least 50~ms prior to fluorescence imaging. During this time, the bottom dipole beam, if present, is turned off for 10~ms as a clean-up procedure to remove trapped atoms beyond the tweezer depth-of-field, followed by ramping back on to induce the conveyor transport. Right before fluorescence imaging, the bottom dipole beam power is again ramped off in 2~ms while the tweezer beam power is ramped up to $10~$mW to strengthen the stationary tweezer lattice during imaging.

\section{Fluorescence imaging and filters}
While the same set of MOT/PGC cooling beams can be used for fluorescence imaging in freespace, the scattered photons from the membrane/nanostructure surface contributes to significant background counts during fluorescence imaging. Instead, we adopt additional pairs of linearly polarized beams for imaging, as discussed in the following. These beams are tuned to -20~MHz below cesium $F=4\rightarrow F'=5$ resonance and contain a weak mixture of $F=3\rightarrow F'=4$ repumping component. Imaging beam peak intensity corresponds to around half of the saturation intensity (total beam power $\sim10~\mu$W).

To image atomic fluorescence with single atom sensitivity, it is important to reduce the reflection and scattered tweezer light from the dielectric surface, and the scattered imaging beam photons from the dielectric nanostructures lying within the depth of field of our imaging system. The former can be removed with stacked interference filters. For the latter, neither frequency nor aperture filtering are possible. Nevertheless, our dielectric structures, especially the membrane, are fabricated with low impurity and low surface roughness. As a result, polarization filtering is sufficient to reduce the background counts below single atom fluorescence counts. We insert a combination of quarter and half waveplates and a polarizing cube in the Fourier space of the imaging system (Fig. \ref{figSM_app}) to filter out scattered imaging beam photons with high extinction. Atomic fluorescence on the other hand is unpolarized and can still be imaged with $\sim$ 50~\% reduced counts.
 
For imaging trapped atoms on the membrane, as shown in Figs.~\ref{fig:wg}c,d-\ref{fig:conveyor}, atomic fluorescence images are taken with a beam intersecting the membrane with a 35 degree incidence angle from the bottom side of the membrane (Fig.~\ref{fig1}: IM2). After passing through the membrane, the beam is re-collimated and retro-reflected back to balance the radiation pressure. We have adjusted the imaging beam polarization so that its projection on the membrane surface is parallel to the polarization of the tweezer beam for polarization filtering both beams.
 
Using the same beam path (IM2) to image trapped atoms on a nanostructure, however, will result in higher background fluorescence counts due to photons scattering off the structure. The background counts are comparable to single atom fluorescence counts. To further suppress scattering, in Fig.~\ref{fig1}d we adopt a different imaging beam path that intersects the membrane surface from the top at a shallow 8 degree angle  (Fig.~\ref{fig1}: IM1), allowing us to adjust the beam polarization to be nearly parallel to the optical axis, and thus reducing photon scattering into the objective. At this shallow angle, around 52\% of the beam power is reflected off the membrane. We pick the reflected beam and retro-reflect it back for radiation pressure balancing. The background counts in this beam path is reduced to a similar level as those of Figs.~\ref{fig:wg}c,d. 

\section{Forming the optical conveyor belt} 
The 935~nm light for the tweezer lattice is derived from a Ti:Sapphire laser of a narrow linewidth ($<$100~kHz). The light is split into two parts for the top tweezer beams and the bottom dipole beam, respectively. The tweezer trap is controlled by two acousto-optical deflectors (Fig. \ref{figSM_app}), marked as AOD-x and AOD-y respectively, while the bottom dipole beam is frequency shifted by passing through an acousto-optical modulator (AOM) twice via retro-reflection. The radio frequency sources driving the AODs and the AOM are synchronously generated by an arbitrary waveform generator. The total frequency shift in the tweezer beam of interest is $\nu_\mathrm{x}+\nu_\mathrm{y} \equiv  \nu$, where $\nu_{\mathrm{x(y)}}$ are the radio frequency components in AOD-x(-y), respectively. The AOM is initially driven by a radio frequency $\nu/2$, which is ramped to $\nu/2 + \Delta\nu/2$ during the conveyor transport. This leads to a total frequency shift of $\nu + \Delta \nu$ and a differential shift $\Delta\nu$ between the top tweezer beam and the bottom dipole beam. The accumulated phase shift $\Delta \phi(t) = \int_0^t \Delta \nu(t') dt'$ is used to control the transport distance $\Delta z(t) = \frac{\lambda_\mathrm{t}}{2}\Delta \phi(t)$ \cite{Schrader2001}. For this study, the optical path difference between the two beams is not interferometrically stabilized. We have measured a $\sim$200~Hz phase noise between the two beams, which may cause up to $\pm1~\mu$m uncertainty during the long hold time $\tau=9$~ms for the transport.

\begin{figure}[t]
\includegraphics[width=1\columnwidth]{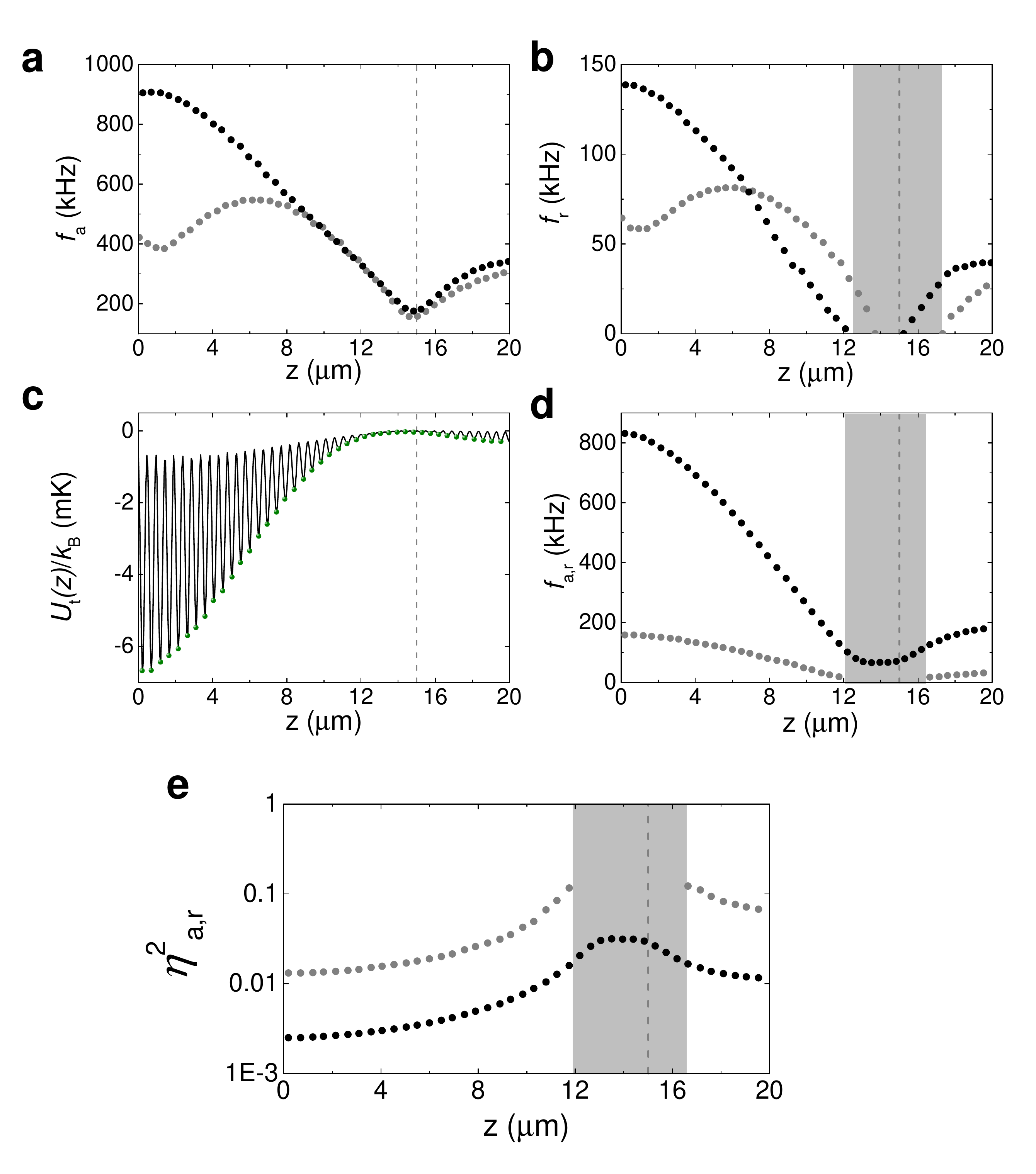}%  
\caption{\textbf{Trap frequencies and Lamb-Dicke parameters in a tweezer lattice on membrane.}  (a-b) Axial and radial trap frequencies of the conveyor belt lattice potential, $f_\mathrm{a}$ and $f_\mathrm{r}$, versus site position $z$ when the bottom dipole beam is in-phase (black circles) and out-of-phase (gray circles), respectively. (c) Potential line-cut $U_\mathrm{t}(z)$ through the center of the stationary tweezer lattice formed by a 10~mW tweezer beam during fluorescence imaging, whose axial (black circles) and radial (gray circles) trap frequencies are plotted in (d). (e) The Lamb-Dicke parameters in the axial (black circles) and radial (gray circles) directions of the stationary tweezer lattice. The vertical dashed lines mark the position of the tweezer depth of field. The gray shaded areas mark the region with negative radial potential curvature and thus with no radial trapping.}
\label{figSMtrapf}
\end{figure}

\section{Calculations of the tweezer lattice and conveyor belt potentials} Tweezer potentials on arbitrary nanostructures, as presented in Fig.~\ref{fig1}, are calculated using a commercial simulator based on the finite-difference time-domain method. Due to the simple planar geometry, the stationary tweezer lattice [Fig.~\ref{figSMtrapf} (c)] and the conveyor belt potentials on the membrane (Fig.~\ref{fig:conveyor}) can be evaluated analytically \cite{novotny2012principles} by considering the angular reflection spectrum from the layered dielectrics, which we compute using the Fresnel equations and a transfer matrix method. The position of the first site in the stationary lattice is determined by the thickness of the layered membrane, and can be tuned to $z\lesssim 100~$nm. For a tweezer beam of wavelength $\lambda_\mathrm{t}=935~$nm and on a membrane of thickness reported in this study, the first site is at $z\sim 200~$nm. 

In Fig.~\ref{fig:conveyor}a, we show that the lattice potential in a conveyor belt is strong everywhere except near the depth-of-field $z_\mathrm{dof} = 2\lambda_t/\mathrm{NA}^2$ where the tweezer intensity on the optical axis vanishes \cite{novotny2012principles}. To further illustrate trap weakening near this point,  in Fig. \ref{figSMtrapf} we evaluate the trap curvature and therefore the trap frequency in both the axial (along the z-axis) and the radial directions (in the x-y plane) at different site locations in the optical conveyor belt. We find that not only the trap weakens, the radial trap curvature actually turns negative beyond $z\approx 12~\mu$m, violating the condition for stable trapping. This generic feature manifests in all tweezer lattice potentials, including the stationary lattices on the waveguide, Fig.~\ref{fig1}, and on the membrane, Fig. \ref{figSMtrapf}.

\section{Analysis of single atom fluorescence and loading statistics} 
To estimate trap loading probability, in the first imaging period, we perform a single Gaussian fit only to the low count region in the histograms (Fig.~\ref{fig:wg}) where the background manifest as a single peak. We then calculate the fitted occurrence for the background with zero atom occupancy to estimate the probability for trapping $n\geq 1$ atoms in the tweezer lattice. This estimation includes contributions from loosely trapped atoms, likely localized away from the tweezer focus, that do not appear in the second imaging period. 

We analyze the loading statistics in the second imaging period for atoms stably trapped within the tweezer lattice. We find that the number-resolved occurrence $C(I)$ of CCD counts $I$ can be empirically fitted by a composite Gaussian model
\begin{eqnarray}
C(I)&=&\frac{1}{\sqrt{\pi}}\left[\frac{P_0}{w_{bg}}e^\frac{-(I-I_{bg})^2}{w_{bg}^2} \right. \nonumber \\ 
&& \left. +\sum_{n=1}^{n_\mathrm{max}} \frac{P_n}{w\sqrt{nI_a+I_{bg}}}e^\frac{-(I-nI_a-I_{bg})^2}{w^2(nI_a+I_{bg})}\right] \nonumber
%C(I)=\frac{1}{\sqrt{\pi}}\left[\frac{P_0}{w_{bg}}e^\frac{-(I-I_{bg})^2}{w_{bg}^2} 
%+\sum_{n=1}^{n_\mathrm{max}} \frac{P_n}{w\sqrt{nI_a+I_{bg}}}e^\frac{-(I-nI_a-I_{bg})^2}{w^2(nI_a+I_{bg})}\right]
\end{eqnarray}
up to $n_\mathrm{max}=3$, where $P_n$ is the occurrence for occupancy of $n=0, 1, 2,3$ atoms, $I_\mathrm{bg}$ is the average background count, $w_\mathrm{bg}$ is the background noise, $I_\mathrm{a}$ is the average single atom fluorescence count, and $w$ is an excess noise factor beyond shot-noise. For imaging on the waveguide structure shown in Fig.~\ref{fig:wg}b, $(I_\mathrm{bg,w},w_\mathrm{bg,w},I_\mathrm{a,w},w_\mathrm{w}) = (370,134,1037,11)$; for imaging on the membrane as shown in Fig.~\ref{fig:wg}d, we obtain $(I_\mathrm{bg,m},w_\mathrm{bg,m},I_\mathrm{a,m},w_\mathrm{m}) = (221,138,853,8.4)$. From the composite Gaussian fit, we then estimate the trap loading probability from the fitted occurrence at each occupancy $n$. The results are plotted in insets of Fig.~\ref{fig:wg}. 

For imaging on the waveguide and the membrane, respectively, we note that their background and single atom fluorescence counts differ due to different imaging conditions such as intensity and polarization orientation. Imaging on the membrane has better signal-to-background ratio ($I_\mathrm{a,m}/I_\mathrm{bg,m}=3.9>I_\mathrm{a,w}/I_\mathrm{bg,w}=2.8$). In both cases, the CCD count noise (width of each Gaussian peak) is much higher than the shot noise due to excess noise from the CCD and the spatial variation of the imaging beam intensity which forms an unstabilized standing-wave pattern scanning across the tweezer trap during imaging.

\section{Heating during fluorescence imaging}
The fluorescence imaging is taken with only a pair of counter propagating beams (either IM1 or IM2 in Fig.~\ref{fig1}). As a result, we expect transverse heating due to photon scattering. In the study, we determine that $\sim$1000 averaged CCD counts are recorded for single atom fluorescence. Using this number, we estimate that each atom scatters $N_\mathrm{p}\sim$45,000 photons during the 30~ms imaging time, after taking into account the CCD electronic settings such as A/D converter efficiency ($3e^-$ per count), electron-multiplier gain ($G=30$), and quantum efficiency ($QE\sim 0.5$), as well as the total transmittance $T\sim15~$\% of the optical system and finally the $\Omega\approx3~$\% objective collection efficiency. Without further cooling assisted by trap mixing or trap suppression in transverse heating, an atom may be heated up by $\sim \frac{2}{3}N_\mathrm{p}E_\mathrm{R}/k_\mathrm{B}\approx 3~$mK, gaining enough kinetic energy to leave the trap. Here $E_\mathrm{R} = h^2/2\lambda_\mathrm{a}^2m$ is the photon recoil energy, $h$ is the Planck constant, and $m$ is the cesium atomic mass. 

Tight confinement in the tweezer lattice provides significant suppression of trap excitation during imaging. We evaluate the Lamb-Dicke parameters $\eta_\mathrm{a,r}^2 = E_\mathrm{R}/hf_\mathrm{a,r}$ for the stationary tweezer lattice [Fig. \ref{figSMtrapf} (e)], which should provide an estimate of the suppression factor of recoil heating (near the trap ground state) during imaging. It is clear that losing radial confinement becomes the major limiting factor for those trapped away from the tweezer focus. In Fig. \ref{figSMtrapf} (e), $\eta_\mathrm{r}^2$ rises up quickly approaching $z\approx12~\mu$m. This could qualitatively explain why a quick reduction of fluorescence counts is observed for $\Delta z_\mathrm{f}>0$ in Fig.~\ref{fig:conveyor}d. It is also suggestive that all atoms observed during the second imaging period in Fig.~\ref{fig:wg} should be trapped and imaged well within $z<10~\mu$m.

\begin{figure}[t]
\includegraphics[width=1\columnwidth]{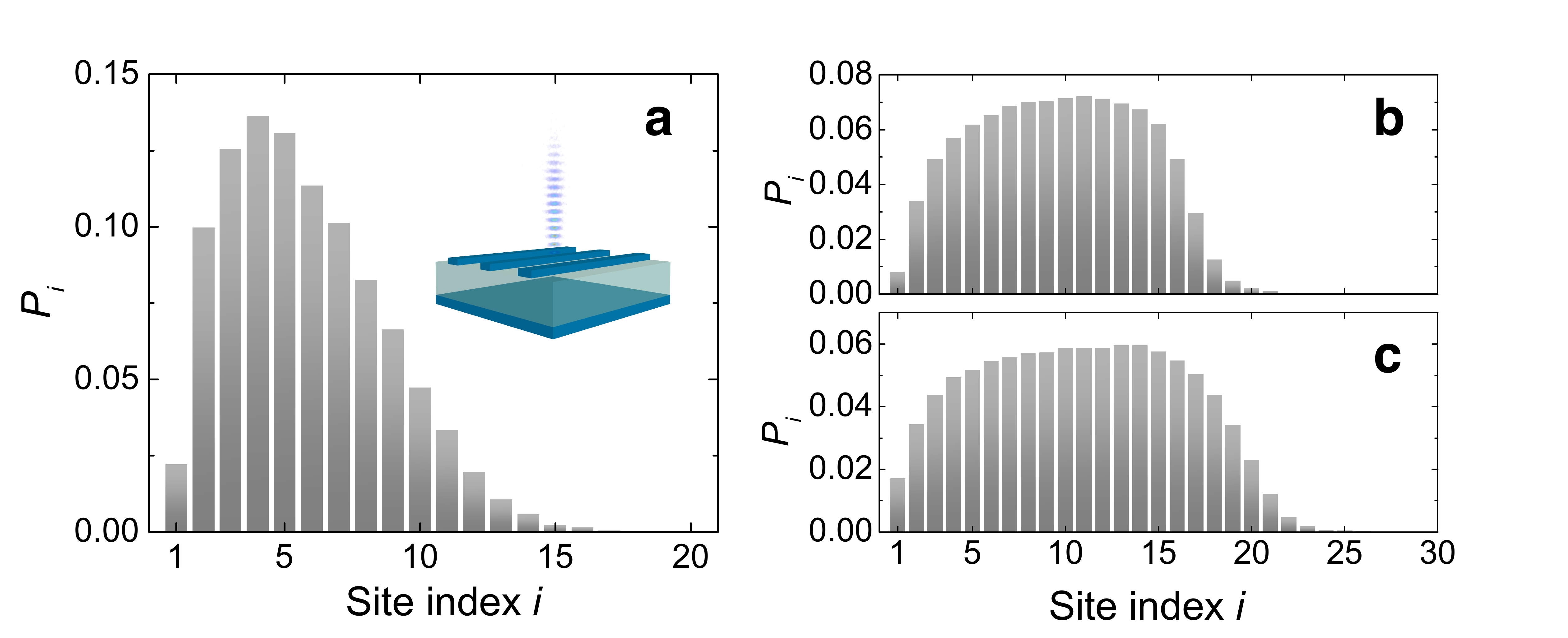}%  
\caption{\textbf{Monte Carlo simulation of trap loading probability in a tweezer lattice.} (a) Individual site loading probability on a waveguide as in Fig. 1. Inset depicts trap probability density above the waveguide. (b-c) Individual site loading probability in a conveyor belt on a membrane when the bottom dipole beam is in the in-phase (b) and the out-of-phase (c) conditions, respectively.}
\label{figMC}
\end{figure}

\section{Monte Carlo simulation of tweezer lattice loading}
To estimate trap loading efficiency and probability distribution within a tweezer lattice, we perform Monte Carlo simulations of Doppler cooling in a $(10~\mu$m$)^2\times20~\mu$m~(L$\times$W$\times$H) region with corresponding lattice potentials above the photonic structures as shown in Figs.~\ref{fig1} and \ref{fig:conveyor}. Actual loading efficiency with PGC may differ from this calculation. We include the effect of atom-surface interactions by approximating the surface Casimir-Polder (CP) potential with $U_\mathrm{cp}(z) = -C_4/z^3[z+\lambdabar]$, where $z$ is the atom-surface distance, $C_4/h= 158 (267)~$Hz$\cdot\mu$m$^4$ on \siot (\sitnf), $h$ is the Planck constant, and $\lambdabar = 136~$nm \cite{stern_simulations_2011}. To estimate loading efficiency in each tweezer lattice, 10$^6$ trajectories are calculated, each beginning with an atom randomly entering from the top or the four sides of the simulation boundaries with a velocity sampled from a thermal ensemble of temperature $T=$20$~\mu$K. Following 1~ms of loading simulation, the number of trajectories remaining in the trap region are counted to estimate the trap efficiency and their final positions are recorded for calculating the trap probability distribution within the tweezer lattice.

For a lattice on the waveguide as shown in Fig.~\ref{fig1}, our simulation results indicate that around $P_\mathrm{tot,w}=1.3$\% of total trajectories can be loaded into the tweezer lattice. Figure \ref{figMC} plots the probability distribution evaluated from those bound trajectories, indicating that trapped atoms are well localized within $\sim10$ lattice sites or $5~\mu$m range of the tweezer focus.

For a lattice on the membrane as shown in Fig.~\ref{fig:conveyor}, we simulate the condition when the bottom dipole beam is in-phase (out-of-phase) with respect to the surface reflection of the tweezer beam, giving a much larger trap probability $P_\mathrm{tot,m}=27 $\%(13\%) due to larger trap volume offered by the bottom beam and overall deeper lattice depths (Fig.~\ref{fig:conveyor}). Site loading probability distributions, as shown in Fig.~\ref{figMC} (b-c), are more or less uniform within the first 20 sites in the simulation region.

In all cases in Fig.~\ref{figMC}, site loading probabilities in the evanescent wave region ($z < \lambda_\mathrm{a}$) of the photonic structures are relatively low. At most 2\% can be found in the first site closest to the dielectric surface. This inefficiency is predominantly due to atom-surface CP interactions and the presence of other trap sites that reduces the solid angle for entering the first site near the surface. Other trap ramping strategies may be devised to increase loading probability at the first site.

The total loading efficiency obtained from the MC results is in qualitative agreement with some of our experimental observations. In $t=$10~ms experiment loading time, the estimated number of atoms traversing a surface area $A\approx O(100\mu$m$^2$) is $N_\mathrm{a}=\rho_0 A \bar{v} t \approx 100~$atoms, where we have used $\bar{v} = 3~$cm s$^{-1}$. The estimated number of trappable atoms is $\bar{n}_\mathrm{MC,w} = N_\mathrm{a}P_\mathrm{tot,w}\approx 1.3$ for the waveguide (Fig.~\ref{fig1}) and $\bar{n}_\mathrm{MC,m}\approx20$ for the tweezer lattice on membrane. While the former is close enough to experimental estimate $\bar{n}_\mathrm{wg} \gtrsim 1$ for trapped atoms on the structure [Fig.~\ref{fig:wg}a], the latter $\bar{n}_\mathrm{MC,m}\gg \bar{n}_\mathrm{lattice}=3.6$ for trapping with a bottom dipole beam [Fig.~\ref{fig:conveyor}d]. In fact, $\bar{n}_\mathrm{MC,m}$ is already close to the number of populated lattice sites ($\sim$20) enclosed in the region. This strongly suggests that collisional blockade should manifest during trap loading, which is not included in the MC calculation and will be addressed in future studies. 

\section{Fitting the fluorescence data following the conveyor transport}
We make a remark that defocusing effect cannot explain the counts in Fig.~\ref{fig:conveyor}d. For the sake of understanding, we first discuss ideal imaging without heating atoms out of the trap. We model the atomic fluorescence as emission from symmetric point dipole sources. With our objective $\mathrm{NA}=0.35$ and assuming diffraction limited imaging, paraxial point-spread function remains a fairly good approximation \cite{novotny2012principles}. In Fig.~\ref{fig:conveyor}c, we count the fluorescence within an area $\mathcal{A} = 6\times6$~CCD pixels. We calculate the expected total counts from a defocused atom within this area $\mathcal{A}$. The estimated count reduction in Fig.~\ref{fig:conveyor}d assumes atoms are initially randomly distributed within the tweezer lattice and are being transported out of the tweezer focus. We also consider contributions from defocused `image atoms' due to membrane reflection. However, the reflectance $R(\theta)\approx 0.3$ is small for $\theta<\theta_\mathrm{max}=\sin^{-1}\mathrm{NA}=20^\circ$ and cannot be fully responsible for the reduced counts.

We justify that, at $\Delta z_\mathrm{f}>0$, atoms either escape the trap during transport or being heated or pushed out of the trap during imaging. We thus empirically fit the data with a simple exponential, giving a function $I(z)$ of single atom fluorescence counts versus distance from the membrane surface; $I(z)=0$ for $z\leq0$. For $\Delta z_\mathrm{f} < 0$, we model fluorescence counts as $I(z_i+\Delta z_\mathrm{f})$ for an atom initially trapped in a lattice site at $z=z_i$ and later being transported by a distance $\Delta z_\mathrm{f}$. Using this empirical model, we perform least-squared fitting to the downward-conveying data by assuming a Poisson average of $\bar{n}_\mathrm{lattice}$ trapped atoms randomly distributed along the tweezer lattice within $0<z_i \leq z_\mathrm{max}$ followed by transport $\Delta z_\mathrm{f}$. Each `fit' is an average of 100 random trap configurations and the gray regions accounts for the error of the mean. Figure~\ref{fig:conveyor}d shows the fit result with only two fit parameters, giving $\bar{n}_\mathrm{lattice} = 3.6$ and $z_\mathrm{max} = 10.3~\mu$m, corresponding to a maximum site index $i_\mathrm{max}=22$ in reasonable agreement with the MC simulation results shown in Fig.~\ref{figMC}.

\end{document}